
\input harvmac
\noblackbox

\Title{\vbox{\baselineskip12pt\hbox{hep-th/0312104}
\hbox{SU-ITP-03/33, SLAC-PUB-10259, TIFR/TH/03-26} }}
 {\vbox{ {\centerline{Flux Compactifications on}
\smallskip
\centerline{Calabi-Yau Threefolds}} } }

\centerline{Alexander
Giryavets$^a$\footnote{*}{On leave from Steklov Mathematical Institute, Moscow, Russia}\footnote{$^1$}{giryav@stanford.edu}, Shamit
Kachru$^a$\footnote{$^2$}{skachru@stanford.edu},
Prasanta K. Tripathy$^b$\footnote{$^3$}{prasanta@theory.tifr.res.in} and Sandip P. Trivedi$^b$\footnote{$^4$}{sandip@tifr.res.in} }
\smallskip
\centerline{$^{a}$ Department of Physics and SLAC}
\centerline{Stanford University} \centerline{Stanford, CA
94305/94309 USA}
\smallskip
\centerline{$^{b}$ Tata Institute for Fundamental Research}
\centerline{Homi Bhabha Road, Mumbai 400 005, INDIA}

\medskip

\noindent The presence of RR and NS three-form fluxes in type IIB
string compactification on a Calabi-Yau orientifold gives rise to
a nontrivial superpotential $W$ for the dilaton and complex
structure moduli.  This superpotential is computable in terms of
the period integrals of the Calabi-Yau manifold.  In this paper,
we present explicit examples of both supersymmetric and
nonsupersymmetric solutions to the resulting 4d ${\cal N}=1$
supersymmetric no-scale supergravity, including some
nonsupersymmetric solutions with relatively small values of $W$.
Our examples arise on orientifolds of the hypersurfaces in
$WP^{4}_{1,1,1,1,4}$ and $WP^{4}_{1,1,2,2,6}$.  They serve as
explicit illustrations of several of the ingredients which have
played a role in the recent proposals for constructing de Sitter
vacua of string theory.

\Date{December 2003}

\noblackbox

\def\IP{\relax{\rm I\kern-.18em P}}

\lref\specialg{A. Strominger, ``Special Geometry,'' Comm. Math.
Phys. {\bf 133} (1990) 163\semi P. Candelas and X. de la Ossa,
``Moduli Space of Calabi-Yau Manifolds,'' Nucl. Phys. {\bf B355}
(1991) 455.}

\lref\afont{A. Font, ``Periods and Duality Symmetries in
Calabi-Yau Compactifications,'' Nucl. Phys. {\bf B391} (1993) 358,
hep-th/9203084.}

\lref\gp{B. Greene and R. Plesser, ``Duality in Calabi-Yau Moduli
Space,'' Nucl. Phys. {\bf B338} (1990) 15.}

\lref\klemmfour{A. Klemm, B. Lian, S. Roan and S.T. Yau,
``Calabi-Yau Fourfolds for M and F-theory Compactifications,''
Nucl.Phys. {\bf B518} (1998) 515, hep-th/9701023.}

\lref\dave{D. Morrison, ``Picard-Fuchs equations and mirror maps
for hypersurfaces,'' hep-th/9111025.}

\lref\CY{P.~Candelas, G.~Horowitz, A.~Strominger, and E.~Witten,
``Vacuum Configurations for Superstrings,'' Nucl. Phys. {\bf B258}
(1985) 46.}

\lref\cdv{P. Candelas, X. de la Ossa and F. Rodriguez Villegas,
``Calabi-Yau Manifolds over Finite Fields, I,'' hep-th/0012233.}

\lref\hkp{J. Hsu, R. Kallosh and S. Prokushkin, ``On Brane
Inflation with Volume Stabilization,'' hep-th/0311077.}

\lref\tye{H. Firouzjahi and S. Tye, ``Closer Towards Inflation in
String Theory,'' hep-th/0312020.}

\lref\SW{A.~Strominger and E.~Witten, ``New Manifolds For
Superstring Compactification,'' Commun.\ Math.\ Phys.\  {\bf 101}
(1985) 341.}

\lref\Polchinski{J.~Polchinski, {\it{String Theory, Volume II:
Superstring Theory and Beyond,}} Cambridge University Press,
1998.}

\lref\Wsb{E.~Witten, ``Symmetry Breaking Patterns In Superstring
Models,'' Nucl.\ Phys.\ B {\bf 258} (1985) 75.}

\lref\hosono{S. Hosono, A. Klemm, S. Theisen and S.T. Yau,
``Mirror Symmetry, Mirror Map, and Applications to Calabi-Yau
Hypersurfaces,'' Comm. Math. Phys. {\bf 167} (1995) 167,
hep-th/9308122.}

\lref\GKP{S. Giddings, S. Kachru, and J. Polchinski, ``Hierarchies
from Fluxes in String Compactifications,'' Phys. Rev. {\bf D66}
(2002) 106006, hep-th/0105097.}

\lref\FP{A. Frey and J. Polchinski, ``{\cal N}=3 Warped
Compactifications,'' Phys. Rev. {\bf D65} (2002) 126009,
hep-th/0201029.}

\lref\freyrev{A. Frey, ``Warped Strings: Selfdual Flux and
Contemporary Compactifications,'' hep-th/0308156.}

\lref\evads{A. Maloney, E. Silverstein and A. Strominger ``de
Sitter Space in Noncritical String Theory,'' hep-th/0205316.}

\lref\Verlinde{H. Verlinde, ``Holography and Compactification,''
Nucl. Phys. {\bf B580} (2000) 264, hep-th/9906182\semi C. Chan, P.
Paul and H. Verlinde, ``A Note on Warped String
Compactification,'' Nucl. Phys. {\bf B581} (2000) 156,
hep-th/0003236.}

\lref\evatong{E. Silverstein and D. Tong, ``Scalar Speed Limits
and Cosmology: Acceleration from D-cceleration,'' hep-th/0310221.}

\lref\ibanez{P. Camara, L.E. Ibanez and A. Uranga, ``Flux-induced
SUSY-breaking soft terms,'' hep-th/0311241\semi M. Grana, ``MSSM
parameters from supergravity backgrounds,'' Phys.Rev. {\bf D67} (2003) 066006,
hep-th/0209200.}

\lref\kklmmt{S. Kachru, R. Kallosh, A. Linde, J. Maldacena, L.
McAllister and S. P. Trivedi, ``Towards Inflation in String
Theory,'' JCAP {\bf 0310} (2003) 013, hep-th/0308055\semi E.
Silverstein and D. Tong, ``Scalar Speed Limits and Cosmology:
Acceleration from D-cceleration,'' hep-th/0310221\semi J. Hsu, R.
Kallosh and S. Prokushkin, ``On Brane Inflation with Volume
Stabilization,'' hep-th/0311077\semi H. Firouzjahi and S.H. Tye,
``Closer Towards Inflation in String Theory,'' hep-th/0312020.}

\lref\landscape{L. Susskind, ``The Anthropic Landscape of String
Theory,'' hep-th/0302219.}

\lref\uranga{ J. Cascales, M. Garcia de Moral, F. Quevedo and A.
Uranga, ``Realistic D-brane Models on Warped Throats: Fluxes,
Hierarchies and Moduli Stabilization,'' hep-th/0312051\semi J.
Cascales and A. Uranga, ``Chiral 4d String Vacua with D-branes and
Moduli Stabilization,'' hep-th/0311250\semi J. Cascales and A.
Uranga, ``Chiral 4d ${\cal N}=1$ String Vacua with D-Branes and
NSNS and RR Fluxes,'' JHEP {\bf 0305} (2003) 011,
hep-th/0303024.}

\lref\ralph{R. Blumenhagen, D. L\"ust and T. Taylor, ``Moduli
Stabilization in Chiral Type IIB Orientifold Models with Fluxes,''
Nucl.Phys. {\bf B663} (2003) 319, hep-th/0303016.}

\lref\KST{S. Kachru,~M. Schulz and S. P. Trivedi,
``Moduli Stabilization from Fluxes in a Simple IIB Orientifold,''
JHEP {\bf 0310} (2003) 007, hep-th/0201028.}

\lref\TT{P. Tripathy and S. P. Trivedi, ``Compactification with Flux
on K3 and Tori,''  JHEP {\bf 0303} (2003) 028, hep-th/0301139.}

\lref\braneinf{G. Dvali and S.H. Tye, ``Brane Inflation,'' Phys.
Lett. {\bf B450} (1999) 72, hep-th/9812483\semi S. Alexander,
``Inflation from D-Anti-D-Brane Annihilation,'' Phys. Rev. {\bf
D65} (2002) 023507, hep-th/0105032\semi G. Dvali, Q. Shafi and M.
Solganik, ``D-brane Inflation,'' hep-th/0105203\semi C.P. Burgess,
M. Majumdar, D. Nolte, F. Quevedo, G. Rajesh and R. Zhang, ``The
Inflationary Brane-Antibrane Universe,'' JHEP {\bf 07} (2001) 047,
hep-th/0105204\semi G. Shiu and S.H. Tye, ``Some Aspects of Brane
Inflation,'' Phys. Lett. {\bf B516} (2001) 421, hep-th/0106274.}

\lref\TV{T.~R.~Taylor and C.~Vafa, ``RR flux on Calabi-Yau and
partial supersymmetry breaking,'' Phys.\ Lett.\  {\bf B474} (2000)
130, hep-th/9912152\semi
P. Mayr, ``On Supersymmetry Breaking in String Theory and its
Realization in Brane Worlds,'' Nucl. Phys. {\bf B593} (2001) 99,
hep-th/0003198.}

\lref\kv{S. Kachru and C. Vafa, ``Exact Results for ${\cal N}=2$
Compactifications of Heterotic Strings,'' Nucl. Phys. {\bf B450}
(1995) 69, hep-th/9505105.}

\lref\GVW{S.~Gukov, C.~Vafa, and E.~Witten, ``CFT's From
Calabi-Yau Four-folds,'' Nucl.\ Phys.\ {\bf B584} (2000) 69,
hep-th/9906070.}

\lref\cklt{ G.~Curio, A.~Klemm, D.~L\"ust and S.~Theisen, ``On the
Vacuum Structure of Type II String Compactifications on Calabi-Yau
Spaces with H-Fluxes,'' Nucl.Phys. {\bf B609} (2001) 3,
hep-th/0012213.}

\lref\cogp{ P. Candelas, X. de la Ossa, P. Green and L. Parkes
``A Pair of Calabi-Yau Manifolds as an exactly
soluble superconformal theory,'' Nucl. Phys. {\bf B359} (1991) 21 .}

\lref\bcofhjq{ P. Berglund, P. Candelas, X. de la Ossa, A. Font,
T. Hubsch, D. Jancic and F. Quevedo,
``Periods for Calabi-Yau and Landau-Ginzburg Vacua,''
Nucl.Phys. {\bf B419} (1994) 352,  hep-th/9308005.}

\lref\cofkm{ P. Candelas, X. de la Ossa, A. Font,
S. Katz and D. Morrison,
``Mirror Symmetry for Two Parameter Models - I,''
Nucl.Phys. {\bf B416} (1994) 481, hep-th/9308083.}

\lref\Sen{A. Sen, ``Orientifold limit of F-theory vacua,'' Phys.
Rev. {\bf D55} (1997) 7345, hep-th/9702165.}

\lref\kt{ A.~Klemm and S.~Theisen,
``Considerations of One-Modulus Calabi-Yau Compactifications:
Picard-Fuchs Equations, K\"ahler Potentials and
Mirror Maps,'' Nucl.Phys. {\bf B389} (1993) 153, hep-th/9205041.}

\lref\kklmv{ S.~Kachru, A.~Klemm, W.~Lerche, P.~Mayr and C.~Vafa
``Nonperturbative Results on the Point Particle Limit
of N=2 Heterotic String Compactifications,''  Nucl.Phys. {\bf B459} (1996) 537,
hep-th/9508155.}

\lref\maa{ G.~Moore, ``Attractors and Arithmetic,'' hep-th/9807056.}

\lref\maaa{ G.~Moore, ``Arithmetic and Attractors,'' hep-th/9807087.}

\lref\kklt{S. Kachru, R. Kallosh, A. Linde and S. P. Trivedi, ``de
Sitter Vacua in String Theory,'' Phys.Rev. {\bf D68} (2003) 046005,
hep-th/0301240.}

\lref\douglas{S. Ashok and M. Douglas, ``Counting Flux Vacua,''
hep-th/0307049; M. Douglas, ``The Statistics of String/M-theory
Vacua,'' JHEP {\bf 0305} (2003) 046, hep-th/0303194.}

\lref\quevedo{C. Escoda, M. Gomez-Reino and F. Quevedo,
``Saltatory de Sitter String Vacua,'' hep-th/0307160\semi C.
Burgess, R. Kallosh and F. Quevedo, ``de Sitter string vacua from
supersymmetric D-terms,'' JHEP {\bf 0310} (2003) 056, hep-th/0309187.}

\lref\oldflux{J. Polchinski and A. Strominger, ``New vacua for
type II string theory,'' Phys. Lett. {\bf B388} (1996) 736,
hep-th/9510227\semi K. Becker and M. Becker, ``M-theory on eight
manifolds,'' Nucl. Phys. {\bf B477} (1996) 155,
hep-th/9605053\semi J.Michelson, ``Compactifications of type IIB
strings to four-dimensions with non-trivial classical potential,''
Nucl. Phys. {\bf B495} (1997) 127, hep-th/9610151\semi
K. Becker and M. Becker, ``Supersymmetry Breaking, M Theory and
Fluxes,'' JHEP {\bf 010} (2001) 038, hep-th/0107044\semi
M. Haack and J. Louis, ``M theory compactified on Calabi-Yau
fourfolds with background flux,'' Phys. Lett. {\bf B507} (2001)
296, hep-th/0103068\semi J. Louis and A. Micu, ``Type II theories
compactified on Calabi-Yau threefolds in the presence of
background fluxes,'' Nucl.Phys. {\bf B635} (2002) 395, hep-th/0202168.}

\lref\oldwarp{K. Dasgupta, G. Rajesh and S. Sethi, ``M theory,
orientifolds and G-flux,'' JHEP {\bf 9908} (1999) 023, hep-th/9908088\semi
B. Greene, K. Schalm and G. Shiu, ``Warped compactifications in
M and F theory,'' Nucl. Phys. {\bf B584} (2000) 480, hep-th/0004103.}

\lref\chaud{
M. Bianchi, G. Pradisi and A. Sagnotti, Nucl. Phys. {\bf B376} (1991) 365\semi
S. Chaudhuri, G. Hockney and J. Lykken, ``Maximally Supersymmetric
String Theories in $D<10$,'' Phys. Rev. Lett. {\bf 75} (1995) 2264,
hep-th/9505054\semi
S. Chaudhuri and D. Lowe, ``Type IIA-Heterotic Duals with Maximal
Supersymmetry,'' Nucl. Phys. {\bf B459} (1996) 113, hep-th/9508144\semi
S. Chaudhuri and D. Lowe, ``Monstrous String Duality,'' Nucl. Phys.
{\bf B469} (1996) 21, hep-th/9512226.}

\lref\svw{S. Sethi, C. Vafa and E. Witten, ``Constraints on
Low Dimensional String Compactifications,'' Nucl. Phys. {\bf B480} (1996)
213, hep-th/9606122.}

\lref\KS{I. Klebanov and M. Strassler, ``Supergravity and a Confining
Gauge Theory: Duality Cascades and $\chi SB$ Resolution of Naked
Singularities,'' JHEP {\bf 0008} (2000) 052, hep-th/0007191.}

\lref\ferrara{R. D'Auria, S. Ferrara and S. Vaula, ``${\cal N}=4$
gauged supergravity and a IIB orientifold with fluxes,'' New J.
Phys. {\bf 4} (2002) 71, hep-th/0206241\semi S. Ferrara and M.
Porrati, ``${\cal N}=1$ no-scale supergravity from IIB
orientifolds,'' Phys. Lett. {\bf B545} (2002) 411,
hep-th/0207135\semi R. D'Auria, S. Ferrara, M. Lledo and S. Vaula,
``No-scale ${\cal N}=4$ supergravity coupled to Yang-Mills: the
scalar potential and super-Higgs effect,'' Phys. Lett. {\bf B557}
(2003) 278, hep-th/0211027\semi R. D'Auria, S. Ferrara, F.
Gargiulo, M. Trigiante and S. Vaula, ``${\cal N}=4$ supergravity
Lagrangian for type IIB on $T^6/Z_2$ orientifold in presence of
fluxes and D3 branes,'' JHEP {\bf 0306} (2003) 045,
hep-th/0303049\semi L. Andrianopoli, S. Ferrara and M. Trigiante,
``Fluxes, supersymmetry breaking and gauged supergravity,''
hep-th/0307139\semi B. de Wit, Henning Samtleben and M. Trigiante,
``Maximal Supergravity from IIB Flux Compactifications,''
hep-th/0311224.}

\lref\susskind{L. Susskind, ``The anthropic landscape of string
theory,'' hep-th/0302219.}

\font\cmss=cmss10 \font\cmsss=cmss10 at 7pt

\def\IC{\relax\hbox{$\inbar\kern-.3em{\rm C}$}}
\def\IR{\relax{\rm I\kern-.18em R}}
\def\Z{\relax\ifmmode\mathchoice
{\hbox{\cmss Z\kern-.4em Z}}{\hbox{\cmss Z\kern-.4em Z}}
{\lower.9pt\hbox{\cmsss Z\kern-.4em Z}}
{\lower1.2pt\hbox{\cmsss Z\kern-.4em Z}}\else{\cmss Z\kern-.4em
Z}\fi}

\def\fto{\tilde{f}_1}
\def\ftt{\tilde{f}_2}

\def\hto{\tilde{h}_1}

\newsec{Introduction}

Given the vast array of possible string compactifications to 4d,
it is very useful to find large classes of constructions which can
be studied systematically.  One of the most interesting questions
regards the detailed structure of the potential for the plethora
of moduli fields that typically arise. In the most familiar case
of Calabi-Yau compactifications, these moduli include the complex
structure and K\"ahler moduli of the Calabi-Yau space, and the
dilaton or string coupling constant. Knowledge of the potential
for these moduli is crucial in making concrete models of particle
physics, in designing cosmological scenarios in string theory, and
in understanding what if anything string theory says about the
cosmological constant problem.

It has been realized over the past several years that in fact in
generic compactifications of string theory to four dimensions, one
is allowed  to turn on fluxes of some of the p-form RR and NS
fields in the compact dimensions (see e.g. {\refs{\oldflux \GVW
\oldwarp \cklt \GKP \KST \FP {--} \TT}}; dual descriptions of some
very simple flux compactifications appear in \chaud). We shall
focus on the specific case of the type IIB theory on a Calabi-Yau
orientifold, which preserves 4d ${\cal N}=1$ supersymmetry at the
KK scale. The relevant background fluxes are those of the RR
three-form field strength $F_{(3)}$ and the NS three-form field
strength $H_{(3)}$. Given a choice of these fluxes, i.e. of two
integral three-forms obeying a tadpole condition determined by the
precise orientifold, one can compute the superpotential $W$ very
explicitly in terms of periods of the holomorphic three-form
$\Omega$ \refs{\GVW,\TV}. An appropriate framework for analyzing
these solutions in some detail was developed in \GKP, and explicit
examples involving tori and K3 surfaces were studied in detail in
\refs{\KST \FP {--} \TT}. The possibility of constructing models
with significant warping was described in \refs{\Verlinde,\GKP}
(the construction of \GKP\ follows closely the work of \KS; see
also \oldwarp\ for earlier papers about constructing warped flux
compactifications). Recently, models with chiral low-energy gauge
theories were discussed in roughly this framework
\refs{\uranga,\ralph}, and a catalogue of flux-induced soft susy
breaking terms on D3-branes was derived \ibanez.  An up to date
review of this subject can be found in \freyrev, and work
developing the relevant gauged supergravities to describe this
class of compactifications can be found in \ferrara.

Given the rather explicit form of the Gukov-Vafa-Witten
superpotential that controls much of the dynamics of the moduli in
these compactifications, it is reasonable to hope that one can
understand the properties of the solutions (at least in the
leading-order no-scale supergravity approximation) rather
explicitly. However, to date, the only (compact) examples
presented in complete detail have involved toroidal orientifolds
or Calabi-Yau spaces with reduced holonomy. Here, we present some
explicit solutions of the IIB flux equations for orientifolds of
``generic'' Calabi-Yau threefolds, whose holonomy fills out
$SU(3)$.  This is of more than academic interest: such examples
are closely related to some proposals for constructing de Sitter
vacua in string theory \refs{\kklt,\quevedo}, and for more
precisely estimating the number of metastable string vacua
\douglas.

We will find two surprises in our analysis.  First, we will find
that supersymmetric solutions of the flux equations do exist.
Given an elementary counting argument which we will review below,
this is by itself somewhat surprising.  Perhaps more importantly,
we will find that simple nonsupersymmetric solutions to the flux
equations (still at vanishing potential $V=0$ in the no-scale
approximation, as described in \GKP) with small values of $W$ also
exist. This is a bit surprising given the small numbers of fluxes
we will be turning on.  These examples provide support for the
assertion in e.g. \kklt\ that by discretely tuning the choice of
fluxes in manifolds with large $b_3$, one can attain small values
of $W$.\foot{Of course this is a meaningful notion only after one
has fixed the K\"ahler invariance, otherwise one should specify
something physical like the gravitino mass. We will describe the
conventions in which we desire ``small $W$'' below, they coincide
with those in \kklt.}

The organization of this paper is as follows. In \S2, we describe
the basic facts about the two models (which we call model A and
model B) that we will be studying -- the threefold geometries, the
relevant orientifold actions, and the lift to an F-theory
description.  We also describe the special (small) subclass of
fluxes we will be turning on, and the symmetries of the resulting
potential which guarantee that we can consistently solve the
equations with many of the CY moduli frozen at a special symmetric
locus.  This saves us from having to solve the Picard-Fuchs
equations for hundreds of independent periods in the two models.
In \S3, we give a more precise formulation of the problems of
interest, and we present details about the period integrals in the
two models. In \S4, we give examples of supersymmetric solutions
in model B. In \S5, we give examples of nonsupersymmetric
solutions in both models, including some with small $W$. We close
with a discussion in \S6. In two appendices, we include more
details about various computations in the two models.

\newsec{The two models of interest}
\subsec{The Calabi-Yau threefolds}

 We will be studying orientifolds
of two different Calabi-Yau threefolds.  Model A will be
constructed starting with the threefold $M_A$ which arises as a
hypersurface in $WP^4_{1,1,1,1,4}$

\eqn\modela{4 x_{0}^2 + x_{1}^8 + x_{2}^8 + x_{3}^8 + x_{4}^8 - 8
\psi x_0 x_1 x_2 x_3 x_4 ~=~0.} This threefold has $h^{1,1}=1$ and
$h^{2,1} = 149$. It served as one of the first examples of mirror
symmetry \refs{\dave \kt {--} \afont}, generalizing the seminal work of
\cogp\ on the quintic. After taking the quotient by the maximal
group of scaling symmetries as in the Greene-Plesser construction
of mirror manifolds \gp, the modulus $\psi$ describes the single
complex structure modulus of a mirror manifold $W_A$. Then, the
classical geometry of complex structure deformations of $W_A$
reproduces the quantum K\"ahler moduli space of $M_A$. However we
will be interested not in the mirror, but in $M_A$ itself. In this
context, there are many other terms that could appear deforming
the complex structure in \modela; we explain why it will be
consistent to neglect these deformations in \S2.3. We will also
describe the production of an appropriate orientifold in the next
subsection.

Our second model, model B, is based on a Calabi-Yau threefold with
$h^{1,1}(M_B) = 2$ and $h^{2,1}(M_B) = 128$. The manifold arises
as a (resolution of) a hypersurface in $WP^4_{1,1,2,2,6}$

\eqn\modelb{x_0^2 + x_1^{12} + x_2^{12} + x_3^6 + x_4^6 - 12 \psi
x_0 x_1 x_2 x_3 x_4 - 2\phi x_1^6 x_2^6 ~=~0.} It was also studied
as one of the first examples of mirror symmetry in two-parameter
models (its mirror $W_B$ has a two-parameter complex structure
moduli space), in \refs{\cofkm,\hosono}. In addition it played a
role as one of the first examples of ${\cal N}=2$ heterotic/type
II string duality \refs{\kv,\kklmv}. The restriction to the two
moduli $\phi$ and $\psi$ is natural in the Greene-Plesser
construction of mirror symmetry, where they parametrize the
subspace of the moduli space of $M_B$ which is invariant under the
maximal group of scaling symmetries.  Again, since we will be
interested in $M_B$ and not in its mirror, we could in principle
add many additional terms to \modelb; we shall explain their
absence below in \S2.3.

\subsec{The orientifolds}

We are interested in ${\cal N}=1$ compactifications of the type
IIB theory on $M_A$ and $M_B$.  To break the symmetry from ${\cal
N}=2$ to ${\cal N}=1$, we must orientifold.  The orientifolds we
study will fall in the class described in \GKP, and can in fact be
produced by Sen's construction \Sen\ which relates Calabi-Yau
fourfold compactifications of F-theory to IIB orientifolds. In
fact, the fastest way for us to compute the relevant properties of
the orientifolds will be to follow Sen's procedure, and specify
the F-theory fourfolds.

For model A, consider the fourfold $X_A$ given by the Calabi-Yau
hypersurface in $WP^{5}_{1,1,1,1,8,12}$.  This is model 5 in Table
B.4 of \klemmfour. Following the procedure of \Sen, one
immediately sees that it reduces to an orientifold of $M_A$ in an
appropriate limit. It has $\chi(X_A) = 23,328$, which means that
in the IIB picture there will be a tadpole condition \svw

\eqn\tada{N_{D3} + N_{\rm flux} = {\chi(X_A)\over 24} = 972.} Here,
$N_{D3}$ is the number of space-filling D3 branes one chooses to
insert, and $N_{\rm flux}$ is the D3 brane charge carried by the $H_{(3)}$
and $F_{(3)}$ fluxes.

For model B, the fourfold $X_B$ is given by the Calabi-Yau
hypersurface in $WP^5_{1,1,2,2,12,18}$, model 21 in Table B.4 of
\klemmfour. Since $\chi(X_B) = 19728$, there will be 822 units of
D3 brane charge to play with in this model.  Again following \Sen,
one sees that in an appropriate limit, it becomes an orientifold
of type IIB on $M_B$.

What the observations of this subsection teach us is that
appropriate ${\cal N}=1$ orientifolds of $M_A$ and $M_B$ do exist,
with specified (rather large!) amounts of D3 brane charge that
must be inserted (via fluxes or space-filling branes) to satisfy
the tadpole condition.  In fact, the Sen construction is
consistent with producing orientifolds on the loci of complex
structure moduli space specified in \modela\ and \modelb.  To see
this, one simply observes that the Sen orientifold action amounts
to taking $x_0 \to -x_0$, composed with worldsheet orientation
reversal. It may be confusing that the monomial $x_0 x_1 x_2 x_3
x_4$ appears in \modela\ and \modelb\ (since it is not invariant).
However using the ring relations (setting the partial derivatives
of the defining equation to zero), this can be re-expressed in
terms of $(x_1 x_2 x_3 x_4)^2$ with an extra factor of $\psi$, and
the deformation is manifestly invariant. More explicitly, in model
A for example, one can define a new coordinate ${\tilde x}_0= x_0
- 4 \psi x_1 x_2 x_3 x_4$ and express the defining equation
\modela\ in terms of this variable. Then only $\psi^2$ will appear
in the defining polynomial, and the orientifold action will
identify ${\tilde x}_0 \rightarrow -{\tilde x}_0$.
  In the presentation given,
$\psi$ appears, while in the manifestly invariant prescription
only $\psi^2$ appears. There are identifications on the $\psi$
moduli space that mean that $\psi \to -\psi$ is a symmetry in both
models (we will say more about this when we discuss the periods),
and in the presentation of the manifolds in \modela\ and \modelb,
one should take Sen's prescription to also act with the modular
symmetry $\psi \to -\psi$.  The reader who finds this confusing is
advised to think instead in the picture with the deformation
parametrized by $\psi^2 (x_1 x_2 x_3 x_4)^2$.

One last comment.
The three-form fluxes we turn on are  allowed by the orientifold symmetry.
The NS and RR two forms,  $B_{2}, C_{2}$, have odd intrinsic  parity under
the orientifold symmetry, and the relevant  three-cycles are also odd \foot{
One quick way to see this is as follows. In the definitions of \kt, \cofkm, \bcofhjq,
 $\Omega$, the holomorphic
three form, is even  under the orientifold symmetry. From, eq. (3.10), (3.19), we see that
 the periods are odd, since $\psi \rightarrow -\psi$. This shows that the three-cycles
are odd.}.
Thus turning on the three forms $H_{3}, F_{3}$, along these three cycles is allowed.

\subsec{Special loci and symmetries}

In both model A and model B, there are many complex structure
deformations (even including only those which are preserved by the
orientifold action). Turning on arbitrary fluxes, the calculation
of the flux superpotential would then require a solution of the
Picard-Fuchs equations for a vast number of periods. We are not
going to proceed in this manner.

Instead, we make the following simple observation.  The special
families of defining equations we have written down in \modela\
and \modelb\ are invariant under large groups of global
symmetries.  The symmetry group is ${\cal G}_A = \Z_2 \times
\Z_8^2$ for model A and ${\cal G}_B = \Z_2 \times \Z_6^2$ for
model B. All deformations which we have not included explicitly in
\modela\ and \modelb\ transform nontrivially under ${\cal
G}_{A,B}$.  As argued in e.g. \S3\ of \cdv\ (where the example of
the quintic is discussed in detail), this means that the
Picard-Fuchs equations simplify greatly, if one is interested only
in a subset of the periods.  Namely, in model A, there are four
periods which will coincide with those of $W_A$, and in which the
other moduli of $M_A$ (which do not appear in \modela) can only
appear with high enough powers to maintain ${\cal G}_A$
invariance. Similarly, in model B, there are six periods which
will coincide with those of $W_B$, and in which the other moduli
of $M_B$ will only appear with high enough powers to maintain
${\cal G}_B$ invariance.

Roughly speaking, what happens in model A is that there is a
four-dimensional subspace of $H_3(M_A)$ which is the homology dual
to the four dimensional subspace in $H^3(M_A)$ spanned by the
(2,1) form associated to $\psi$, it's dual (1,2) form, and the
(3,0) and (0,3) forms.  One can compute the periods in this
subspace of $H_3(M_A)$ and they depend only on higher powers of
the deformations absent in \modela, due to ${\cal G}_A$
invariance.  Similar remarks apply to model B, where however there
are six relevant periods instead of four (since one has two
deformations, $\phi$ and $\psi$).

The physical interpretation of this is clear.  Suppose we turn on
only fluxes consistent with the ${\cal G}_{A,B}$ symmetries. Then
certain terms (low order terms in the ${\cal G}$-charged moduli)
are forbidden from appearing in the flux superpotential. The
moduli which appear only at higher order in $W$ can be
consistently set to zero (as we have done in the defining
equations \modela\ and \modelb) because of the symmetry. They will
generically be constrained by a higher order potential, which is
guaranteed to vanish at their origin.  Since this only holds if we
turn on a restricted set of fluxes which maintain the ${\cal G}$
invariance, we can choose fluxes only through four three-cycles in
model A and six in model B.  These are simply related to the
cycles which appear in computing the periods of the mirror
manifolds $W_A$ and $W_B$.\foot{One difference one must be careful
to account for involves factors of $\vert {\cal G}\vert$ in the
proper normalization of the periods over integral cycles.}

For these reasons, in further discussion of the models, we shall
always set the complex structure moduli except $\psi$ in model A
and $(\phi,\psi)$ in model B to be at their origin (where the
${\cal G}$ symmetries are unbroken).  We shall also neglect the
dependence of the periods on these moduli, since as we have
explained, it is of high enough order that the equations for these
other moduli cannot obstruct solutions on the symmetric locus.

\newsec{Detailed structure of the models}
\subsec{Basic facts common to both models}

\noindent{\it{Homology and cohomology bases}}

We will work with a symplectic homology basis for the subspaces of
$H_3$ of interest to us. The basis of three-cycles $A_a$ and $B^a$
($a=1,2$ for model A and $a=1,2,3$ for model B) and the basis for
integral cohomology $\alpha_a$ and $\beta^a$ satisfy
\eqn\homcohbasis{ \int_{A_a}\alpha_{b}=\delta^{a}_{b},~~~~~~
\int_{B^b}\beta^{a}=-\delta^{a}_{b},~~~~~~
\int_{M}\alpha_{a}\wedge\beta^{b}=\delta^{b}_{a}.} The holomorphic
three form can be represented in terms of periods in this basis as
follows: \eqn\holform{ \int_{A_a}\Omega=z^a,~~~~~~
\int_{B^a}\Omega={\cal G}_a,~~~~~~ \Omega=z^{a}\alpha_a-{\cal
G}_{a}\beta^{a}.} In addition \eqn\ombom{
\int_{M}\Omega\wedge\bar{\Omega}=\bar{z}^a{\cal G}_{a}
-z^a\bar{{\cal G}}_{a}= \bar{z}^a{\partial {\cal G} \over \partial
z^a} -z^a{\partial \overline{{\cal G}} \over \partial \bar{z}^a}
=-\Pi^{\dagger}\cdot\Sigma\cdot \Pi~.}  Here, we have introduced
the prepotential ${\cal G}(z^1,z^2)$, the period vector $\Pi$
(whose entries are the periods \holform), and the matrix
\eqn\sigma{ \Sigma=\pmatrix{ 0 & 1\cr -1 & 0 \cr}} whose entries
are two by two matrices.  This structure is common to all
Calabi-Yau compactifications, see e.g. \specialg.

\smallskip
\noindent{\it{Fluxes, Superpotential and K\"ahler Potential}}

The NS and RR fluxes admit the following quantization condition
\eqn\quant{ F_{(3)}=(2\pi)^2\alpha'( f_a [B_a]+f_{a+k}
[A_a]),~~~~~~~ H_{(3)}=(2\pi)^2\alpha'( h_a [B_a]+h_{a+k} [A_a]) }
with integer $f_i$ and $h_i$.  Here $k=2$ for model A and $k=3$
for model B, and $a$ runs over $1,2$ for model A and $1,2,3$ for
model B. Here we also used the notation $[A_a]=\alpha_a$ and
$[B_a]=\beta_{a}$. Using this notation, we  find the following
expression for $N_{\rm flux}$ \eqn\nflux{ N_{\rm flux}={1\over
(2\pi)^4(\alpha')^2}\int_{M} H_{(3)}\wedge
F_{(3)}=f^{T}\cdot\Sigma\cdot h.} The superpotential is given by
\eqn\superp{ W=\int_{M} (F_{(3)}-\tau H_{(3)})\wedge \Omega=
(2\pi)^2\alpha' (f\cdot \Pi-\tau h\cdot \Pi).} The K\"ahler
potential for the dilaton-axion and complex moduli is given by
\eqn\kahler{ K=K_{\tau}+K_{\rm
c.m.}=-\ln(-i(\tau-\bar{\tau}))-\ln(i\int_{M}\Omega\wedge
\bar{\Omega})= -\ln(-i(\tau-\bar{\tau}))-\ln( - i
\Pi^{\dagger}\cdot\Sigma\cdot \Pi).} where c.m. is $\psi$ for
model A and $(\psi,\phi)$ for model B. For more discussion of the
low-energy effective action of these IIB orientifolds, see e.g.
\GKP.

\smallskip
\noindent{\it{Conditions for solutions}}

The supersymmetry conditions for flux vacua are given by
\eqn\supw{
W=0,~~~~~~D_{\tau}W=0,~~~~~~e^{K}G^{a\bar{b}}D_{a}W\overline{D_{b}W}=0,
} where for model A $a,b=\psi$, and for model B they run over
$\phi, \psi$. We have kept the $e^K$ in \supw\ because given the
conventions for normalizing $\Omega$ in e.g. \refs{\kt,\cofkm},
this factor can sometimes make a difference.

If one wishes to find no-scale vacua without supersymmetry, the
conditions \supw\ are relaxed; one need not impose $W=0$.  In such
solutions the potential still vanishes at tree-level, but there
are non-vanishing F-auxiliary VEVs for some K\"ahler moduli \GKP.

\subsec{Periods for Model A}

In \kt\ (following \cogp), the relevant periods for model A are
given as follows.  In the Picard-Fuchs basis, the fundamental
period is given by

\eqn\wo{ w_{0}(\psi)=(2\pi i)^3{\pi \over 8}
\sum_{n=1}^{\infty}{1\over\Gamma(n) \prod_{i=0}^{4}\Gamma(1-{n
\over 8}\nu_{i})} {\exp(i{\pi \over 8}7n) \over \sin({\pi n \over
8})}(4 \psi)^{n} ~.} This expression is valid for $\vert\psi\vert
< 1$.  Here the $\nu_i$ are the weights of the $WP^4$, i.e.
$1,1,1,1,4$. We will choose another gauge and normalization in
comparison with \refs{\cogp,\kt}. The gauge of \refs{\cogp,\kt} is
convenient for considering the fundamental period in the vicinity
of $\psi=\infty$. For the case $\psi=0$ at hand, it is useful to
make a gauge transformation of the holomorphic three form and
corresponding transformation of the K\"ahler potential
\eqn\gaugepsi{ \Omega(\psi)\rightarrow{1\over
\psi}\,\Omega(\psi),~~~~~~ K_{\psi}\rightarrow
K_{\psi}+\ln|\psi|^2. } Also, since we are interested in the
orientifold of $M_A$, not in the mirror, the normalization of the
fundamental period differs by $|{\cal G}_A|$ in comparison with
\refs{\cogp,\kt}. This just follows from the definition of the
fundamental period, which is given by an integral on the cycle
$|x_i|=\delta$ for $i=0,...,4$ and small $\delta$.

In terms of the fundamental period, a basis for the
periods is given by
\eqn\pfbas{w^T_{A} ~=~{1\over \psi} (w_0(\alpha^2 \psi),w_0(\alpha
\psi),w_0(\psi),w_0(\alpha^7 \psi)),}
where $\alpha = \exp({\pi i \over 4})$.

Now, we are really interested in the periods in a symplectic
basis.  The periods in the symplectic basis $\Pi^T_A = ({\cal
G}_1, {\cal G}_2, z^1, z^2)$ can be expressed in terms of those in
the Picard-Fuchs basis by means of a linear transformation:

\eqn\lintr{\Pi_A ~=~m_A \cdot w_A,} where the matrix $m_A$ is given by

\eqn\misa{m_A =\pmatrix{-{1 \over 2} & -{1 \over 2} & {1 \over 2}
& {1 \over 2}\cr 0 & 0 & -1 & 0\cr -1 & 0 & 3 & 2 \cr 0 & 1 & -1 &
0\cr}.}

It follows from these formulae that in the vicinity of $\psi=0$,
we can expand the period vector as
\eqn\piais{\Pi_A  =
c_0 p_0 + c_2 p_2 \psi^2 + c_4 p_4 \psi^4 + \cdots}
Here
the vectors $p_k$ are given by \eqn\pktilde{p_k = ~m_A \cdot \tilde{p}_k}
with
\eqn\pvecis{\tilde{p}_k^T ~= (\alpha^{2(k+1)},
\alpha^{(k+1)}, 1, \alpha^{7(k+1)}),} and the constants $c_k$ are as follows
\eqn\cis{\eqalign{& c_0 = (2\pi i)^3 {\sqrt{\pi} \over 2 \Gamma^4(7/8)}
{\exp({7 \pi i\over 8
          })\over \sin({\pi \over 8})}, \cr
& c_2 = -(2\pi i)^3 {2\sqrt{\pi} \over  \Gamma^4(5/8)} {\exp({5 \pi i\over 8
          })\over \sin({3\pi \over 8})}, \cr
& c_4 = (2\pi i)^3 {4\sqrt{\pi} \over  \Gamma^4(3/8)} {\exp({3 \pi i \over 8
          })\over \sin({5\pi \over 8})}~.}}
We kept the fourth order terms here in part to show that they are
small compared to the zeroth and second order terms as long as
$\vert \psi \vert \ll 1$, but they will also play an important
role in our nonsupersymmetric solutions in \S5. The solutions we
will present there, will be valid for small $\vert \psi \vert$
(where, in our examples, the $\psi$ modulus will be stabilized in
a self-consistent approximation).

\subsec{Periods for Model B}

In \refs{\bcofhjq,\cofkm} the following power series expansions
for the periods of this threefold are given.  Define a fundamental
period \eqn\wotp{ w_{0}(\psi,\phi)=(2\pi i)^3{1\over 6}
\sum_{n=1}^{\infty}{(-1)^{n}\Gamma({n\over 6})\over\Gamma(n)
\Gamma^{2}(1-{n \over 6})\Gamma(1-{n \over 2})} (12\psi)^{n}
u_{-{n\over 6}}(\phi),~~~~~~\left|{864\psi^6\over
\phi\pm1}\right|<1 ~.} Here the $u_{v}$s are functions of $\phi$
which are given in e.g. \cofkm.  As with model A, we perform a
gauge transformation rescaling the holomorphic three-form by
${1\over \psi}$, and redefine the fundamental period by
multiplying it by $|{\cal G}_B|$. Then in \refs{\bcofhjq,\cofkm}
they find a six dimensional basis for the periods, given by
\eqn\pervecttp{w_B^{T}={1\over
\psi}(w_0(\psi,\phi),w_0(\alpha\psi,-\phi),
w_0(\alpha^2\psi,\phi),w_0(\alpha^{3}\psi,-\phi),
w_0(\alpha^{4}\psi,\phi),w_0(\alpha^{5}\psi,-\phi)),} where
$\alpha=\exp({\pi i \over 6})$.

The periods in a symplectic basis $\Pi^T_{B} = ({\cal G}_1, {\cal G}_2,
{\cal G}_3, z^1, z^2, z^3)$ can be expressed in terms of the
Picard-Fuchs basis of periods $w$ by a linear transformation
$\Pi_B~=~m_B \cdot w_B$, where

\eqn\mtp{ m_{B}=\pmatrix{-1 & 1 & 0 & 0 & 0 & 0\cr {3\over 2} &
{3\over 2} & {1\over 2} & {1\over 2} & -{1\over 2} & -{1\over
2}\cr 1 & 0 & 1 & 0 & 0 & 0 \cr 1 & 0 & 0 & 0 & 0 & 0 \cr -{1\over
2} & 0 & {1 \over 2} & 0 & {1 \over 2} & 0 \cr {1 \over 2} & {1
\over 2} & -{1 \over 2} & {1 \over 2} & -{1 \over 2} & {1 \over
2}\cr}.}

In the vicinity of $\psi=0$ and some regular locus for $\phi$, we
can expand the periods \wotp~ as follows

\eqn\pib{\Pi_B =
c_0 p_0 + c_2 p_2 \psi^2 + \cdots}
where $c_0$ and $c_2$ are given by
\eqn\constB{
c_0=-(2\pi i)^{3}{4\sqrt{\pi}\over \Gamma^{3}({5\over 6})},~~~~~~
c_2=(2\pi i)^{3}(12)^2{1\over 2\pi},
}
and the vectors $p_0$ and $p_2$ are given by
\eqn\pip{\eqalign{&
p_0=m_B\cdot(u_{-{1\over 6}}(\phi)\tilde{p}_{01}
+\alpha u_{-{1\over 6}}(-\phi)\tilde{p}_{02})
\equiv u_{-{1\over 6}}(\phi)p_{01}
+\alpha u_{-{1\over 6}}(-\phi)p_{02}, \cr
& p_2=m_{B}\cdot(u_{-{1\over 2}}(\phi)\tilde{p}_{21}
+i u_{-{1\over 2}}(-\phi)\tilde{p}_{22})
\equiv u_{-{1\over 2}}(\phi)p_{21}
+i u_{-{1\over 2}}(-\phi)p_{22}, \cr
}}
where

\eqn\pippp{\eqalign{
& \tilde{p}^T_{01}=(1,0,\alpha^2,0,\alpha^4,0),~~~~~~
\tilde{p}^T_{02}=(0,1,0,\alpha^2,0,\alpha^4),\cr
& \tilde{p}^T_{21}=(1,0,-1,0,1,0),~~~~~~~~
\tilde{p}^T_{22}=(0,1,0,-1,0,1).\cr
}}

This completes our discussion of the periods for the two models.
We will refer back to the formulae from these sections as the need
arises, when specifying our solutions.

\newsec{Supersymmetric Solutions in Model B}

In this section we will study supersymmetric solutions  in model
B.

\noindent{\it{Supersymmetry conditions for Model B}}

The K\"ahler potential and metric have the
following behavior for small $\psi$ at regular points in the
$\phi$ moduli space
 \eqn\supdetailstpa{ K(\psi,\phi)\sim 1,
~~~~~~ G_{\psi\bar{\psi}}\sim \psi\overline{\psi}, ~~~~~~
G_{\psi\bar{\phi}}\sim\psi\overline{\psi}^2, ~~~~~~
G_{\phi\bar{\psi}}\sim\overline{\psi}\psi^2, ~~~~~~
G_{\phi\bar{\phi}}\sim 1. }
This means that terms with mixed $\psi$
and $\phi$ derivatives do not appear in the potential.

Henceforth the supersymmetry conditions for flux vacua \supw~ are
\eqn\supcondB{
W=0,~~~~~~\partial_{\tau}W=0,~~~~~~{1\over\psi}\partial_{\psi}W=0,
~~~~~~\partial_{\phi}W=0.}

This gives the following supersymmetry conditions \eqn\supfhtpa{
f\cdot  p_{01}=0,~~~~
f\cdot  p_{02}=0,~~~~
h\cdot  p_{01}=0,~~~~
h\cdot  p_{02}=0,~~~~
(f-\tau h)\cdot p_2=0~. }

\smallskip
\noindent{\it{Explicit solutions}}

The first four conditions of \supfhtpa\ are satisfied for rank two
degenerate families of vectors
\eqn\fhtpa{ f=\pmatrix{-2f_2 \cr f_2 \cr f_3 \cr -4f_2-2f_3 \cr
-f_3 \cr 0},~~~~~~~~ h=\pmatrix{-2h_2 \cr h_2 \cr h_3 \cr
-4h_2-2h_3 \cr -h_3 \cr 0}. }
For these families $N_{\rm flux}$ \nflux~ is given by
\eqn\nfluxfhtpa{N_{\rm flux}=3(f_2h_3-h_2f_3)~.}

The last condition of \supfhtpa\ fixes the dilaton-axion to the
following value
\eqn\dilaxtpa{ \tau(\phi)={ u_{-{1\over 2}}(\phi)(f_2+f_3)+i
u_{-{1\over 2}}(-\phi)f_2 \over u_{-{1\over 2}}(\phi)(h_2+h_3)+i
u_{-{1\over 2}}(-\phi)h_2 },} where the $u_{-{1\over 2}}(\phi)$
function (given in \cofkm) is \eqn\unu{ u_{-{1\over
2}}(\phi)={1\over \sqrt{2}\pi}\int_{-1}^{1}d\zeta
{1\over\sqrt{(1-\zeta^2)(\phi-\zeta)}}. } In this model we have a
moduli space of supersymmetric vacua parametrized by $\phi$, with
singularities at complex codimension one (for instance on the
locus $\phi^2 = 1$).

\newsec{Nonsupersymmetric Solutions}

\subsec{Model A}

We  obtain solutions  to the classical supergravity equations for
model A in this section. These solutions break supersymmetry, but
the scale of supersymmetry breaking is somewhat small compared to
the string scale.

The essential idea behind finding these solutions is the following.
We will work in the vicinity of the $\psi=0$ point in moduli space, eq. \wo .
It will turn out that obtaining a supersymmetric solution at $\psi=0$
requires that the ratio of
two fluxes is an irrational number. This condition
cannot be met since the fluxes are quantized to take
integer values. However, it is well known that an irrational number can be arbitrarily
well approximated by
a rational $p/q$. So by discretely tuning the fluxes we will obtain
approximately supersymmetric
 solutions in the vicinity of $\psi=0$.

We expect a similar strategy will be more widely useful in the vicinity of other points in moduli space
and also for other Calabi-Yau compactifications. In the present example, given the restriction on the
total flux which can be turned on, \nflux, the flux integers $p,q$ cannot be taken to be very big,
and one can do only moderately well in lowering the susy breaking
scale. In  other cases where the   total value of flux can be larger, one would expect that the flux
integers  can be made bigger and the approximation to the irrational number can be quite good,
resulting in a small scale of supersymmetry breaking.  Perhaps more importantly,
for simplicity we have  turned on fluxes along only four three-cycles in this analysis.
When more fluxes are turned on one would expect to do better in terms of lowering the
supersymmetry breaking scale.

The analysis below  proceeds in three steps. We  first examine the
requirements for a supersymmetric solution at $\psi=0$.  We then
consider the supersymmetry conditions up to ${\cal O}(\psi^2)$ and
show that for appropriately chosen fluxes they can be met.
Finally, we consider the analysis to higher orders in $\psi$ and
show that the solution breaks supersymmetry at order ${\cal
O}(\psi^4)$.

\noindent{\it{I) Conditions for SUSY Solution at $\psi=0$}}

As discussed previously, the fluxes can be expanded in an integral
cohomology basis

\eqn\fluxbas{F_{(3)} = (2\pi)^2 \alpha^{\prime}(f_a [B_a] +
f_{a+2} [A_a]),~~~~~~ H_{(3)} = (2\pi)^2 \alpha^{\prime}(h_a [B_a]
+ h_{a+2} [A_a])~.}  The superpotential then becomes
\eqn\superais{W_A = f \cdot \Pi_A - \tau h \cdot \Pi_A~.} (We are
neglecting a factor of $(2\pi)^2 \alpha^{\prime}$ in the
normalisation for $W_A$ for now, \superp, this will be restored
towards the end of the section when we calculate the scale of
supersymmetry breaking). The susy conditions provided by \supw\
are \eqn\susyconda{f \cdot \Pi_A = 0,~~~~~~h \cdot \Pi_A = 0,} and
\eqn\susycondb{{1\over\psi}(f - \tau h) \cdot \partial_{\psi}
\Pi_A = 0~.} The last equation, \susycondb, should be understood
as a limiting value  at $\psi=0$. As we will see later on in this
section, in the vicinity of $\psi=0$, the metric $G_{\psi{\bar
\psi}}\sim \psi {\bar \psi}$. Eq. \susycondb\ then follows from
\supw.

Keeping terms up to ${\cal O}(\psi^2)$ we find
\eqn\aseqns{\eqalign{&
 c_0 \tilde{h}\cdot \tilde{p}_0 + c_2 \psi^2 \tilde{h}\cdot \tilde{p}_2 = 0 , \cr
 & c_0 \tilde{f}\cdot \tilde{p}_0 + c_2 \psi^2 \tilde{f}\cdot \tilde{p}_2 = 0 , \cr
 & \tilde{f}\cdot \tilde{p}_2 - \tau \tilde{h}\cdot \tilde{p}_2 = 0.}}
Here for convenience, we redefined the vectors so that
\eqn\redef{\tilde f = f\cdot m_A,~~~~~~\tilde h = h\cdot m_A,}
with the integral flux vectors given by
\eqn\deff{f=(f_1, f_2, f_3, f_4 ),} \eqn\defh{h=(h_1, h_2, h_3,
h_4 ).}
For use below we note that ${\tilde f}$ and ${\tilde h}$ are given in terms of
$ f$ and $h$ as
\eqn\ftilde{\eqalign{
{\tilde f}^T
= \left(\matrix{& \tilde f_1 \cr & \tilde f_2 \cr
& \tilde f_3 \cr & \tilde f_4}
\right)
 = {1\over 2} \left( \matrix{ & -  f_1 - 2 f_3 \cr
               & - f_1 + 2 f_4 \cr
               & f_1 -2 f_2 + 6 f_3 - 2 f_4 \cr
               &  f_1 + 4 f_3}
\right),
}}

\eqn\htilde{\eqalign{
{\tilde h}^T
= \left(\matrix{& \tilde h_1 \cr & \tilde h_2 \cr
& \tilde h_3 \cr & \tilde h_4}
\right)
= {1\over 2} \left( \matrix{ & -  h_1 - 2 h_3 \cr
               & - h_1 + 2 h_4 \cr
               & h_1 -2 h_2 + 6 h_3 - 2 h_4 \cr
               &  h_1 + 4 h_3}
\right).
}}
Also, the total contribution to the D3 charge tadpole from the
fluxes is given by
\eqn\atad{\eqalign{&N_{\rm flux} = f\cdot\Sigma\cdot h^T=
2(\tilde f_4 \tilde h_2 - \tilde f_2 \tilde h_4 + \tilde f_3
\tilde h_1 - \tilde f_1 \tilde h_3) \cr &~~~~~~~~~~+ \tilde f_1 (\tilde h_2
+ \tilde h_4) + \tilde f_2 (\tilde h_3 - \tilde h_1)
    + \tilde f_3 (\tilde h_4 - \tilde h_2) - \tilde f_4 (\tilde h_1 + \tilde
    h_3).}}
Since  $f_i$ and $h_i$ must be integer we note that  $\tilde f_i$'s  and $\tilde h_i$'s are
 rational numbers in general.

Now we are ready to consider the requirements that need to be met for a  susy solution at $\psi=0$.
This imposes two conditions on the flux
\eqn\conda{{\tilde f} \cdot {\tilde p_0}=0,}
and
\eqn\condb{{\tilde h} \cdot {\tilde p_0}=0.}
The dilaton-axion is then given by
\eqn\dila{\tau = {\tilde f \cdot \tilde p_2 \over \tilde h
\cdot \tilde p_2}~.}

Using \pvecis, \conda\ takes the form
\eqn\fombca{i {\tilde f_1} + \alpha {\tilde f_2}+ {\tilde f}_3 -i \alpha  {\tilde f}_4=0.}
To simplify the analysis we  will consider from here on fluxes which meet the condition
\eqn\condc{{\tilde f}_4=0,~~~~~~ {\tilde f_3}={\tilde f_1}.}
Now \conda\ becomes
\eqn\condd{{{\tilde f_1} \over {\tilde f_2}} =-{1 \over \sqrt{2}}.}
As noted above ${\tilde f}_i$ must be rational,  so \condd\ cannot be met.

Similarly, \condb\ takes the form
\eqn\fombd{i{\tilde h_1}+\alpha {\tilde h_2} + {\tilde h_3} -i \alpha {\tilde  h_4} =0.}
Again for easy of analysis we  consider the case where
\eqn\conde{{\tilde h}_2=0,~~~~~~ {\tilde h}_3=-{\tilde h}_1.}
Then \condb\ becomes
\eqn\condf{{{\tilde h_1} \over {\tilde h_4}} ={1 \over \sqrt{2}}.}
This condition again can not be met.

Thus, we cannot have a supersymmetric solution at $\psi=0$ in this model.

\noindent{\it{II) A SUSY solution to ${\cal O}(\psi^2)$}}

We now show that to ${\cal O}(\psi^2)$ the SUSY conditions can be
met in the vicinity of the origin, by appropriately choosing
fluxes. The SUSY conditions \aseqns\ can be solved for $\psi$ and
$\tau$ to get
\eqn\susya{\psi^2 = - {c_0\over c_2} {\tilde f \cdot \tilde p_0 \over
\tilde f \cdot \tilde p_2}~,~~~~~~ \tau = {\tilde f \cdot \tilde p_2 \over \tilde h
\cdot \tilde p_2} ~.}
They also impose restrictions on the fluxes
\eqn\afluxres{(\tilde f \cdot \tilde p_2) (\tilde h \cdot \tilde p_0) - (\tilde
f \cdot \tilde p_0) (\tilde h \cdot \tilde p_2) = 0 ~.}

A straightforward calculation shows that the conditions on fluxes
\afluxres~ can be rewritten as
\eqn\fcond{\eqalign{
&\tilde f_1 \tilde h_3 - \tilde f_3 \tilde h_1 + \tilde
f_4 \tilde h_2 - \tilde f_2 \tilde h_4 = 0,\cr
& \tilde f_1 (\tilde h_4 - \tilde h_2) + \tilde f_2
(\tilde h_1 + \tilde h_3) - \tilde f_3 (\tilde h_2 + \tilde h_4) +
\tilde f_4 (\tilde h_3 - \tilde h_1) ~=~0~.}}
For ease of analysis we will continue to consider fluxes which meet the conditions
\condc\ and \conde. Equation \fcond\ then gives the condition
\eqn\hconda{{\tilde h}_4=-{2 {\tilde h}_1 {\tilde f}_1 \over {\tilde f}_2}.}
We will furnish concrete examples below to show that \condc, \conde\
and \hconda\ can be satisfied for
appropriate integer quantized fluxes.

Once the restrictions on the flux are met, a  solution exists to
this order. Using \condc\ and \conde\ we see that $\psi$ and
$\tau$ are given by \eqn\valex{\psi^2=-i {c_0 \over c_2} \left [{
{\tilde f}_1 + {1 \over \sqrt{2}} {\tilde f}_2 \over {\tilde f}_1
- {1 \over \sqrt{2}} {\tilde f}_2 } \right ],} \eqn\valexb{\tau =
- {i\over \sqrt{2}} {\ftt\over \hto} ~.} For the ${\cal
O}(\psi^2)$ analysis to be valid the resulting value of $\psi$
should satisfy $\vert\psi\vert \ll 1$. We see that this can be
arranged if ${\tilde f}_1 \simeq -{1 \over \sqrt{2}}{\tilde f}_2$,
 as would be expected from our  discussion of a susy vacuum at $\psi=0$
 \foot{Note that if ${\tilde f}_1  \simeq
-{1 \over \sqrt{2}} {\tilde f}_2$, then from \hconda, ${{\tilde h}_1 \over {\tilde h}_4}
 \simeq {1\over \sqrt{2}}$,
so \condf\ is also approximately met.}.

We also note that the total three brane charge carried by the flux, satisfying the conditions
\condc, \conde\ and \hconda, is given by
\eqn\nfluxis{N_{\rm flux} = - 2 {\hto\over \ftt} (\ftt^2 - 4 \fto
\ftt + 2 \fto^2)~.}

As an explicit example we take
\eqn\smwflux{\eqalign{
& f = (-28, 24, 7, -24), \cr
& h = (-34, 41, 12, -17).
}}
The resulting values for ${\tilde f}$ and ${\tilde h}$ are
\eqn\swflxtild{\eqalign{
& \tilde f = (7, -10, 7, 0), \cr
& \tilde h = (5, 0, -5, 7),
}}
 and satisfy the conditions
\condc, \conde\ and \hconda.
Also,  ${ {\tilde f}_1 \over {\tilde f}_2} = -{7 \over 10}$,  so that
$-\sqrt{2}{{\tilde f}_1 \over {\tilde f}_2}  -1 \simeq -0.01$, which is quite small.
As a result  we expect that
$|\psi| \ll 1$ in this example.
Indeed,
inserting the values of $c_0$ and $c_2$ from \S3.2, we obtain from \valex
\eqn\numer{
\psi^2 \simeq 6.47 (1 - i) \times 10^{-3} ~,
}
which is quite small.
The resulting value of the dialton-axion is
$\tau = \sqrt{2} i$.

Also we note that for this example $N_{{\rm flux}}=478$ which is much less than the  maximum allowed
value ${\chi(X_A)/24}=972$.

There is one subtlety  which we have not fully analyzed in this
model. On an orientifold, due to possible ``half cycles" \FP,
sometimes the fluxes $f$ and $h$ need to be even integer (though
often in cases where the subtlety arises, the odd fluxes can be
rendered consistent by turning on fractional fluxes at orientifold
planes \FP). It could be that this subtlety makes the choice of
fluxes \smwflux\ inconsistent.

However one can easily find other examples which involve only even
flux integers. For example one can take
\eqn\moreexp{ f =  (24, -20, -6, 20)~~~~ {\rm and}~~~~ h =
 ( 28 , -34, -10, 14) ~.}
The resulting values for ${\tilde f}$ and ${\tilde h}$ are
\eqn\valftaa{{\tilde f}=(-6,8,-6,0)~~~~ {\rm and}~~~~ {\tilde h}=(-4,0,4,-6).}
Again,  ${ {\tilde f}_1 \over {\tilde f}_2} = -{3\over 4}$  is close to $-{1\over \sqrt{2}}$.
In this case $\psi$ turns out to be
\eqn\numerb{ \psi^2 \simeq 0.038 (i - 1) ~,~~~~~~ \vert\psi\vert^2
\simeq 0.053, }
again much less than one, and the resulting value of the dilaton-axion is $\tau = \sqrt{2} i$.
Also, $N_{\rm flux} = 328$ which is less than  the total allowed value, ${\chi(X_A)/24}=972$,
in this example.

To summarize, we see that appropriate fluxes can be turned on in
model A to meet the conditions of supersymmetry up to ${\cal
O}(\psi^2)$. The resulting vacuum  lies at $\vert\psi\vert \ll 1$,
so we expect the ${\cal O}(\psi^2)$ approximation is good in
determining the location of the vacuum.

\noindent{\it{III) Supersymmetry Breaking at ${\cal O}(\psi^4)$}}

The equations imposing supersymmetry, eq. \supw,  are
overdetermined. Therefore one expects that at higher orders
supersymmetry will be broken in this model. Since $\psi$ is small
in the solution above, and to ${\cal O}(\psi^2)$ we have a SUSY
solution, we expect the resulting supersymmetry breaking to be
somewhat suppressed. By carrying out the analysis to ${\cal
O}(\psi^4)$ in this section we will find this is true. Our
analysis will also ensure that  that the solution found above
extends   in perturbation theory to a solution of the equations of
motion in higher orders.

We will sketch out some of the steps here, more details are
furnished in Appendix A. {}From eq. (A.2) the superpotential is
given by \eqn\superpotf{W_A = c_0 (\tilde f - \tau \tilde h)\cdot
\tilde p_0 + c_2 (\tilde f - \tau \tilde h)\cdot \tilde p_2 \psi^2
+ c_4 (\tilde f - \tau \tilde h)\cdot \tilde p_4 \psi^4 +{\cal
O}(\psi^6).}
A solution  to the classical equations of motion must meet the conditions
\eqn\mina{D_\tau W_A\equiv \partial_\tau W_A  + \partial_\tau K W_A =0,}
and
\eqn\minb{D_\psi W_A \equiv \partial_\psi W_A + \partial_\psi K W_A=0,}
where $K$  is the K\"ahler potential. Eq. (A.6) in Appendix A tells us that  it is
given by
\eqn\xyz{ K = - \log\{- i (\tau - \bar\tau)\} - \log\{  2 [ (2 +
\sqrt{2}) |c_0|^2 + (-2 + \sqrt{2} ) |c_2 \psi^2|^2] + {\cal
O}(\psi^6)\}. }
Eq. \mina\ and \minb\ then take the form
\eqn\expand{\eqalign{
 & D_\psi W_A = 2 \psi \left[
c_2 (\tilde f - \tau \tilde h)\cdot \tilde p_2 + 2 c_4 (\tilde f -
\tau \tilde h)\cdot \tilde p_4 \psi^2 \right. \cr & \left.
~~~~~~~~~ + \left( {2 - \sqrt{2} \over 2 + \sqrt{2}} \right)
\left\{ {\mid c_2\mid \over \mid c_0\mid}\right\}^2 \bar\psi^2
\left(c_0 (\tilde f - \tau \tilde h)\cdot \tilde p_0 + c_2 (\tilde
f - \tau \tilde h)\cdot \tilde p_2 \psi^2 \right) + {\cal
O}(\psi^6) \right], \cr & D_\tau W_A = - \left[ \tilde h\cdot(c_0
\tilde p_0 + \psi^2 c_2 \tilde p_2 + \psi^4 c_4 \tilde p_4)
\right. \cr & \left. ~~~~~~~~~ + {1\over \tau - \bar\tau} \left\{
c_0 (\tilde f - \tau \tilde h)\cdot \tilde p_0 + c_2 (\tilde f -
\tau \tilde h)\cdot \tilde p_2 \psi^2 +
 c_4
(\tilde f - \tau \tilde h)\cdot \tilde p_4 \psi^4 \right\} + {\cal
O}(\psi^6) \right]. }}

Let $\psi_0$ and $\tau_0$ denote  the SUSY preserving solutions
obtained in the previous subsection by working up to ${\cal
O}(\psi^2)$. They satisfy eq. \aseqns . A  consistent solution to
\expand\ can then be obtained by taking $\psi$ and  $\tau$ to be
of the form
\eqn\nowset{\eqalign{ & \psi = \psi_0 + \alpha_\psi \psi_0^3 +
{\cal O}(\psi_0^5), \cr & \tau = \tau_0 + \alpha_\tau \psi_0^2 +
{\cal O}(\psi_0^4). }}

{}From $D_\psi W_A = 0$ we get
\eqn\name{ \psi_0^2\left( -\alpha_\tau c_2 \tilde h \cdot \tilde
p_2 + 2 c_4 (\tilde f - \tau_0 \tilde h)\cdot \tilde p_4 \right) +
{\cal O}(\psi_0^4) = 0, }
solving which we find
\eqn\alphasi{
\alpha_\tau = {2 c_4 (\tilde f - \tau_0 \tilde h)\cdot \tilde p_4 \over c_2 \tilde h \cdot \tilde p_2}.
}
Similarly, $D_\tau W_A = 0$ gives
\eqn\forpsi{ \psi_0^4 \left( 2 c_2 \tilde h \cdot \tilde p_2 ~
\alpha_\psi + c_4 \left\{ \tilde h \cdot \tilde p_4 + {1\over
\tau_0 - \bar\tau_0} (\tilde f - \tau_0 \tilde h) \cdot \tilde p_4
\right\}\right) + {\cal O}(\psi_0^6) = 0, }
and hence
\eqn\alphapsi{
\alpha_\psi = - {c_4 \over 2 c_2 \tilde h \cdot \tilde p_2}
\left\{ \tilde h \cdot \tilde p_4 + {1\over \tau_0 - \bar\tau_0}
(\tilde f - \tau_0 \tilde h) \cdot \tilde p_4 \right\}.
}

Let us now evaluate the ${\cal O}(\psi^4)$ corrections for the two
examples, \smwflux\ and \moreexp, considered above.
Substituting the values of $\psi_0$, $\tau_0$, $\tilde f$, $\tilde h$, $c_i$ and $\tilde p_i$
we find that the resulting values of $\alpha_\psi $ and $\alpha_\tau$ are very close
in the two examples, \smwflux\ and \moreexp. It turns out that  $\alpha_\psi=0$ and
\eqn\forsi{
 \alpha_\tau \simeq 1.073 (1-i).
}

We can now determine the scale of supersymmetry breaking.  The
superpotential is
\eqn\spt{ W_A = c_4 (\tilde f - \tau_0 \tilde h)\cdot \tilde p_4
\psi_0^4 + {\cal O}(\psi_0^6). }
For the choice of fluxes \smwflux, this gives
\eqn\spot{ e^{K/2} |W| \simeq \alpha^{\prime} 4.43 \times 10^{-3}
, }
where we have restored a factor of $(2\pi)^2 \alpha^{\prime}$ in
the relative normalisation between $W$, \supw,  and $W_A$,
\superais. Thus the scale of supersymmetry breaking is indeed
quite   small compared to the string scale.

For the example \moreexp, when all the fluxes are even integers,
we get \eqn\spotb{ e^{K/2} |W| \simeq \alpha^{\prime} 0.125, }
so the scale of breaking is only moderately smaller than the string scale.

Let us end this section with a few more comments. The KKLT
construction \kklt\ involves a small parameter $W_0$. This is the
value of the tree level superpotential in the effective theory
obtained after integrating out the complex structure moduli and
the dilaton.  It is easy to see that  $|W_0|$ in KKLT  is  exactly
the same as $e^{K/2}|W|$ calculated in \spot\ (note for this
purpose that $K$ in \spot\ refers to  the K\"ahler potential of
the complex structure moduli and the dilaton-axion
 fields alone, \xyz,  not the volume modulus).
As was mentioned at the beginning of this section we allowed only four of the fluxes to be turned on
 in model A. It is encouraging to note that even with
this limited number a  modestly   small value of $|W_0|$ has been  obtained in a construction
 which meets several of the other  requirements of the KKLT construction as well.

Finally, we cannot refrain from mentioning one curiosity. For both
examples, \smwflux\ and \moreexp, we have found an alternate
choice of flux which yields a very similar vacuum.
 \smwflux\ is paired with the choice $f=(-20, 17, 5,
-17),\; h=(-48, 58, 17, -24)$, and \moreexp\ with $f=(16, -14, -4,
14),\; h=(40, -48, -14, 20)$. The dilaton
expectation value and $N_{{\rm flux}}$ is  the same in each pair.
The supersymmetry breaking scale is quite similar too,
differing by about one percent in each pair. And the  value of $\psi_0^2$ is
similarly close,  up to a sign.
To the best of our knowledge, this is not the consequence of any known
duality symmetry.

\subsec{Model B}

In this subsection we present some nonsupersymmetric solutions in
model B with nonvanishing $W$ (which is not particularly small).
Let us restrict ourselves to the point $\psi=\phi=0$ and look for
vacua satisfying
\eqn\vacnonb{ W\neq0,~~~~~~D_{\tau}W=0,~~~~~~{1\over\psi}D_{\psi}W=0,
~~~~~~D_{\phi}W=0. }
This gives the following conditions on fluxes
\eqn\vacnonsbf{\eqalign{ &(f-\tau h)\cdot (p_{01}+\alpha
p_{02})\neq0,~~~~~~ (f-\overline{\tau} h)\cdot (p_{01}+\alpha
p_{02})=0,\cr & (f-\tau h)\cdot (p_{21}+ip_{22})=0, ~~~~~~(f-\tau
h)\cdot (p_{01}-\alpha p_{02})=0, }}
where $\alpha=\exp({\pi i\over 6})$. The third condition may be
solved by putting \eqn\flcondthree{ f=\pmatrix{f_1 \cr f_2 \cr f_3
\cr f_4 \cr -4f_1+2f_4 \cr -2f_1-f_2},~~~~~~~~ h=\pmatrix{h_1 \cr
h_2 \cr h_3 \cr h_4 \cr -4h_1+2h_4 \cr -2h_1-h_2}. }
The second and fourth conditions may be used to fix dilaton-axion
to the value \eqn\dilfixnons{ \tau={f\cdot(p_{01}-\alpha
p_{02})\over h\cdot(p_{01}-\alpha p_{02})}.} This solution for
$\tau$ will be consistent with both conditions if
\eqn\cond{ {f\cdot(p_{01}-\alpha p_{02})\over h\cdot(p_{01}-\alpha
p_{02})} ={f\cdot(p^{\dagger}_{01}+\alpha^{\dagger}
p^{\dagger}_{02}) \over h\cdot(p^{\dagger}_{01}+\alpha^{\dagger}
p^{\dagger}_{02})}. }

Now for simplicity, we consider the case when the numerator and
denominator are separately equal. This finally gives the following
two parameter families of fluxes
\eqn\fluxnonsbb{ f=\pmatrix{-2f_2 \cr f_2 \cr -5f_2-2f_4 \cr f_4
\cr 8f_2+2f_4 \cr 3f_2},~~~~~~~~ h=\pmatrix{-2h_2 \cr h_2 \cr
-5h_2-2h_4 \cr h_4 \cr 8h_2+2h_4 \cr 3h_2}. }
The dilaton-axion is equal to
\eqn\fluxnonsbbdil{ \tau={2f_2+f_4+if_2\over 2h_2+h_4+ih_2}.} The
superpotential (evaluated in the vacuum) in this case is equal to
\eqn\fluxnonsbbsup{ W=N_w (f-\tau h)\cdot (p_{01}+ \alpha p_{02})
={i N_w N_{\rm flux}\over 2h_2+h_4+ih_2}, }
where $N_w=(2\pi)^2 c_0 u_{-{1\over 6}}(0)\alpha'$ and $N_{\rm
flux}=6(f_2h_4-h_2f_4)$. It can be expressed in terms of ${\rm
Im}(\tau)$ and the $h$ fluxes
\eqn\supintB{ W=6 i N_w {\rm Im}(\tau) (2h_2+h_4-ih_2). }
Hence,
\eqn\potenB{ e^{K/2}|W|=(2\pi)^2\alpha'\,\sqrt{3\,{\rm
Im}(\tau)}\,|2h_2+h_4-ih_2|. }

For instance, let us choose even fluxes
\eqn\fluxevenBnon{\eqalign{ & f=(8,-4,4,8,-16,-12),\cr &
h=(0,0,4,-2,-4,0). }} This gives $\tau=2i$, $N_{{\rm flux}}=48$,
$h_2=0$, $h_4=-2$ and yields the following \eqn\potenB{
e^{K/2}|W|=(2\pi)^2\alpha^{\prime} 2\sqrt{6}\simeq
\alpha^{\prime}\,193, } which is actually $10^3-10^5$ larger in
comparison with the nonsupersymmetric solutions in model A.

\newsec{Discussion}

In this brief paper, we have seen examples of two interesting
phenomena: the IIB flux equations on some Calabi-Yau threefolds
admit supersymmetric solutions even in the leading approximation
(despite the fact that the no-scale SUSY equations are
overdetermined), and one can find nonsupersymmetric solutions with
relatively small $W$ (even by turning on only a handful of
fluxes). Both of these results provide further motivation to
develop models of particle physics
\refs{\GKP,\uranga,\ralph,\ibanez} and cosmology
\refs{\kklt,\quevedo,\susskind} (including D-brane inflation
\refs{\braneinf,\kklmmt}) in this general framework.

There are a couple of obvious directions for further work.  Our
results have been exploratory in nature, only exhibiting a handful
of solutions in examples which admit easy F-theory lifts.  Any
more general results on the space of solutions in a given example
could complement the ``generic'' analysis of \douglas\ with
detailed specific information, presently only available in the
simple cases of $T^6/Z_2$ and $K3 \times T^2/Z_2$
compactifications.

In addition, the solutions described here provide a further step
towards making completely explicit models of the proposals
\refs{\kklt,\quevedo} for realizing de Sitter vacua in string
theory (see also \evads\ for earlier proposals in noncritical
string theory). Indeed, the F-theory models on $X_A$ and $X_B$
admit stacks of D7 branes which could (when appropriately
stabilized) yield non-Abelian gauge groups and gaugino
condensates.  It is plausible that more work along these lines
could lead to a very explicit realization of the proposal of
\kklt, though of course one would very likely have to turn on more
generic fluxes than the small subset we have used here.

\centerline{\bf{Acknowledgements}}
We would like to thank M. Fabinger, X. Liu, L. McAllister and  E. Silverstein
for interesting discussions on related subjects.
The work of A.G. was partially supported by RFBR grant
02-01-00695 and RFBR grant for young scientists 03-01-06347.
The work of S.K. was supported in part by a David and Lucile
Packard Foundation Fellowship for Science and Engineering, NSF
grant PHY-0097915, and the DOE under contract DE-AC03-76SF00515.
S.P.T. acknowledges support from the Swarnajayanti Fellowship, DST, Govt. of India.
P.T. and S.P.T. also thank the DAE, Govt. of India,   and most of all
the people of India, for  enthusiastically supporting research in String Theory.

\appendix{A}{More Details on Model A}

In this appendix we obtain the nonsupersymmetric solutions for
model A by solving the equations up to ${\cal O}(\psi^4)$.

For convenience, we will first rewrite the superpotential \superais\
\eqn\apaspot{\eqalign{
W_A  = f\cdot \Pi_A - \tau h\cdot \Pi_A
}}
in a different form. Using the Eqs. \redef, \pktilde\ and  \piais,
in the above, we obtain
\eqn\apaspota{\eqalign{
W_A & = \tilde f\cdot m_A^{-1}\cdot \Pi_A
- \tau \tilde h\cdot  m_A^{-1}\cdot \Pi_A \cr
    & = c_0 (\tilde f - \tau \tilde h)\cdot \tilde p_0
        + c_2 (\tilde f - \tau \tilde h)\cdot \tilde p_2 \psi^2 + c_4
(\tilde f - \tau \tilde h)\cdot \tilde p_4 \psi^4 +{\cal
O}(\psi^6). }}

Similarly, using the periods \piais\ the K\"ahler
potential \kahler\ for model A
\eqn\apakahler{
K =  -\ln(-i(\tau-\bar{\tau}))-\ln(- i \Pi_A^{\dagger}\cdot\Sigma\cdot \Pi_A)
}
can be expressed as
\eqn\apakahlerb{\eqalign{
K = & -\ln(-i(\tau-\bar{\tau}))   \cr
& -\ln\left(- i
(c_0 p_0 + c_2 p_2 \psi^2 + c_4 p_4 \psi^4)^{\dagger}\cdot\Sigma\cdot
(c_0 p_0 + c_2 p_2 \psi^2 + c_4 p_4 \psi^4)\right).
}}
It is straightforward to check that
\eqn\checked{\eqalign{
& p_0^{\dagger}\cdot\Sigma\cdot p_2 =
p_0^{\dagger}\cdot\Sigma\cdot p_4 = 0, \cr
& p_0^{\dagger}\cdot\Sigma\cdot p_0 = 2 i (2 + \sqrt{2}), \cr
& p_2^{\dagger}\cdot\Sigma\cdot p_2 = 2 i (- 2 + \sqrt{2}). \cr
}}
Clearly, Eq. \apakahlerb\ becomes
\eqn\apakahlera{ K =  -\ln(-i(\tau-\bar{\tau}))-\ln\{  2 [ (2 +
\sqrt{2}) |c_0|^2 + (-2 + \sqrt{2} ) |c_2 \psi^2|^2] + {\cal
O}(\psi^6)\}. }
Taking the partial derivatives with respect to $\psi$ and $\tau$ we get
\eqn\delpsiofk{\eqalign{ \del_\psi K & =  - { 2 \psi  (-2 +
\sqrt{2} ) |c_2|^2 \bar\psi^2 \over (2 + \sqrt{2}) |c_0|^2 + (-2 +
\sqrt{2} ) |c_2 \psi^2|^2 } \cr & = 2 \psi \left({2 - \sqrt{2}
\over 2 + \sqrt{2}}\right) \left\{ {\mid c_2\mid \over \mid
c_0\mid}\right\}^2 \bar\psi^2 + {\cal O}(\psi^4) ~, }}
and
\eqn\deltauofk{
\del_\tau K = - {1\over \tau - \bar\tau}  ~.
}
We can now evaluate the covariant derivatives \mina\ and \minb\ as
follows: \eqn\minaappa{\eqalign{ D_\psi W_A = & 2 \psi \left[ c_2
(\tilde f - \tau \tilde h)\cdot \tilde p_2 + 2 c_4 (\tilde f -
\tau \tilde h)\cdot \tilde p_4 \psi^2 \right. \cr & \left. +
\left( {2 - \sqrt{2} \over 2 + \sqrt{2}} \right) \left\{ {\mid
c_2\mid \over \mid c_0\mid}\right\}^2 \bar\psi^2 \left(c_0 (\tilde
f - \tau \tilde h)\cdot \tilde p_0 + c_2 (\tilde f - \tau \tilde
h)\cdot \tilde p_2 \psi^2 \right) + {\cal O}(\psi^6) \right] }}
and
\eqn\minbappa{\eqalign{
 D_\tau W_A = & - \left[
\tilde h\cdot(c_0 \tilde p_0 + \psi^2 c_2 \tilde p_2 + \psi^4 c_4
\tilde p_4) \right. \cr & \left. + {1\over \tau - \bar\tau}
\left\{ c_0 (\tilde f - \tau \tilde h)\cdot \tilde p_0 + c_2
(\tilde f - \tau \tilde h)\cdot \tilde p_2 \psi^2 + c_4 (\tilde f
- \tau \tilde h)\cdot \tilde p_4 \psi^4 \right\} + {\cal
O}(\psi^6) \right]. }}

Let $\psi_0$ and $\tau_0$ be the SUSY preserving solutions
obtained from Eq. \aseqns. To find solutions up to ${\cal
O}(\psi^4)$, we observe that the $D_{\tau}W_A = 0 $ condition
implies $\psi \sim \psi_0 + {\cal O}(\psi_0^3)$ and similarly
$D_{\tau}W_A = 0 $ implies $\tau \sim \tau_0 + {\cal
O}(\psi_0^2)$. Thus we take the following ansatz for $\psi$ and
$\tau$:
\eqn\apppsitau{\eqalign{ & \psi = \psi_0 + \alpha_\psi \psi_0^3 +
{\cal O}(\psi_0^5), \cr & \tau = \tau_0 + \alpha_\tau \psi_0^2 +
{\cal O}(\psi_0^4). \cr }}

Putting these in the superpotential \apaspota\ results in
\eqn\resultsw{\eqalign{
W_A & = c_0 (\tilde f - \tau_0 \tilde h)\cdot \tilde p_0
   + c_2 (\tilde f - \tau_0 \tilde h)\cdot \tilde p_2 \psi_0^2
   - \alpha_\tau \tilde h\cdot (c_0 p_0 + \psi_0^2 c_2 p_2) \psi_0^2 \cr
& + 2\alpha_\psi c_2 (\tilde f - \tau_0 \tilde h) \cdot \tilde p_2 \psi_0^4
 + c_4 (\tilde f - \tau_0 \tilde h)\cdot \tilde p_4 \psi_0^4 + {\cal O}(\psi_0^6).
}}
Most of the terms in the r.h.s. of the above equation vanish and
finally $W_A$ becomes
\eqn\finally{ W_A = c_4 (\tilde f - \tau_0 \tilde h)\cdot \tilde
p_4 \psi_0^4 +{\cal O}(\psi_0^6). }
This is expected, since up to quadratic order in $\psi$ both
$W_A(\psi_0,\tau_0)$ and $dW_A(\psi_0,\tau_0)$ vanish. This means
that $W_A(\psi_0,\tau_0) \sim {\cal O}(\psi_0^4)$ as we find in
the above.

Since $\del_\psi K \sim {\cal O}(\psi_0^3)$ and $W_A \sim {\cal
O}(\psi_0^4)$ the $(\del_\psi K) W_A$ term does not contribute
terms up to ${\cal O}(\psi^4)$ in $D_\psi W_A$, whereas $\del_\tau
K \sim 1 + {\cal O}(\psi_0^2)$ and hence both the terms in $D_\tau
W_A$ are significant. Using \apppsitau, we can now easily expand
the r.h.s. of \minaappa\ and \minbappa . Consider first
\eqn\expandrhs{\eqalign{ D_\psi W_A = 2 (\psi_0 + \alpha_\psi
\psi_0^3) \left[ c_2 (\tilde f - \tau_0 \tilde h)\cdot \tilde p_2
- \alpha_\tau \psi_0^2 c_2 \tilde h\cdot \tilde p_2 + 2 c_4
(\tilde f - \tau_0 \tilde h)\cdot \tilde p_4 \psi_0^2 + {\cal
O}(\psi_0^4) \right]. }}
Again, using Eq. \aseqns\ we can simplify it: \eqn\simplifyit{
D_\psi W_A = 2 \psi_0 \left[ - \alpha_\tau \psi_0^2 c_2 \tilde
h\cdot \tilde p_2 + 2 c_4 (\tilde f - \tau_0 \tilde h)\cdot \tilde
p_4 \psi_0^2 + {\cal O}(\psi_0^4) \right]. }
{}From $D_\psi W_A = 0$ we then find
\eqn\dpsiwaapp{ \alpha_\tau = { 2 c_4 (\tilde f - \tau_0 \tilde
h)\cdot \tilde p_4 \over c_2 \tilde h\cdot \tilde p_2} + {\cal
O}(\psi_0^2). }
Similarly, we consider
\eqn\dtauwapp{\eqalign{
D_\tau W_A = - & \left[ \tilde h\cdot(c_0 \tilde p_0 + \psi_0^2 c_2 \tilde p_2
              + 2 \alpha_\psi \psi_0^4 c_2 \tilde p_2
              + \psi_0^4 c_4 \tilde p_4)   \right.  \cr
             & \left. ~~~~+ {1\over \tau_0 - \bar\tau_0}
             \{c_0 (\tilde f - \tau_0 \tilde h)\cdot \tilde p_0
   + c_2 (\tilde f - \tau_0 \tilde h)\cdot \tilde p_2 \psi_0^2
   - \alpha_\tau \tilde h\cdot (c_0 p_0 + \psi_0^2 c_2 p_2) \psi_0^2 \right. \cr
& \left. ~~~~~~~~~~~~~~~~~~~~+ 2\alpha_\psi c_2 (\tilde f - \tau_0 \tilde h) \cdot \tilde p_2 \psi_0^4
 + c_4 (\tilde f - \tau_0 \tilde h)\cdot \tilde p_4 \psi_0^4\}+{\cal O}(\psi_0^6)
\right],
}}
which, upon using Eq. \aseqns, reduces to
\eqn\appdtauw{
D_\tau W_A = - \left[ h\cdot (2 \alpha_\psi c_2 \tilde p_2
              + c_4 \tilde p_4) \psi_0^4
+ {1\over \tau_0-\bar{\tau}_0}c_4 (\tilde f - \tau_0 \tilde
h)\cdot \tilde p_4 \psi_0^4 +{\cal O}(\psi_0^6) \right]. } Solving
$D_\tau W_A = 0$ yields
\eqn\appapsix{
\alpha_\psi = - {c_4 \over 2 c_2 \tilde h \cdot \tilde p_2}
\left\{ \tilde h\cdot \tilde p_4 + {1\over \tau_0 - \bar\tau_0}
(\tilde f - \tau_0 \tilde h)\cdot \tilde p_4
\right\}.
}

\appendix{B}{More Details on Model B}

In this appendix we provide a monodromy group basis for the
hypersurface in $WP^{4}_{1,1,2,2,6}$ in terms of three matrices
denoted by $A$, $T$ and $B$ in the symplectic (large complex
structure) basis and $a$, $t$ and $b$ in the Picard-Fuchs basis.
The former were computed in \kklmv, while the latter appear in a
very similar model in \cofkm.  The two bases are related by the
transformation
\eqn\monrel{ A=m_{B}\cdot a\cdot m^{-1}_{B},~~~~~~ T=m_{B}\cdot
t\cdot m^{-1}_{B},~~~~~~ B=m_{B}\cdot b\cdot m^{-1}_{B},~~~~~~ }
where the matrix $m_{B}$ is defined in \mtp. These monodromies
are obtained by loops in the two parameter moduli space around the
$\Z_{12}$ identified point $\psi=0$, the conifold singularity (which is
$864\psi^6+\phi=\pm 1$), and the strong coupling singularity
($\phi^2=1$).


\subsec{Monodromy group in symplectic (large complex structure)
basis}

In this subsection we reproduce the monodromy matrices given in
\kklmv\ in the symplectic (large complex structure)
basis. They are
\eqn\monodsym{\eqalign{ & A=\pmatrix{-1 & 0 & 1 & -2 & 0 & 0\cr 0
& 1 & 0 & 0 & 2 & 0\cr -1 & 1 & -1 & -1 & 2 & 1 \cr 1 & 0 & 0 & 1
& 0 & 0 \cr -1 & 0 & 0 & -1 & 1 & 1 \cr 1 & 0 & 0 & 1 & 0 &
-1\cr}\cr & T= \pmatrix{1 & 0 & 0 & 0 & 0 & 0\cr 0 & 1 & 0 & 0 & 0
& 0\cr 0 & 0 & 1 & 0 & 0 & 0 \cr -1 & 0 & 0 & 1 & 0 & 0 \cr 0 & 0
& 0 & 0 & 1 & 0 \cr 0 & 0 & 0 & 0 & 0 & 1\cr}\cr & B= \pmatrix{1 &
-1 & 2 & -1 & -2 & 1\cr 0 & 1 & 0 & 2 & 0 & -2\cr 0 & 1 & -1 & 1 &
2 & -1 \cr 0 & 0 & 0 & 1 & 0 & 0 \cr 0 & 0 & 0 & -1 & 1 & 1 \cr 0
& 0 & 0 & 2 & 0 & -1\cr} }}
Here $B=(T_2 A T)^{-1}$, where $T_{2}$ is given in \kklmv.


\subsec{Monodromy group in Picard-Fuchs basis}

In this subsection we compute the monodromy matrices in the
Picard-Fuchs basis explicitly. The monodromy around $\psi=0$ is
the simplest and is given by \eqn\monodromya{
(\psi,\phi)\rightarrow(\alpha\psi,-\phi). } The explicit
expression for the period vector which follows from \wotp\ yields
\eqn\monodpica{ a=\pmatrix{0 & 1 & 0 & 0 & 0 & 0\cr 0 & 0 & 1 & 0
& 0 & 0\cr 0 & 0 & 0 & 1 & 0 & 0 \cr 0 & 0 & 0 & 0 & 1 & 0 \cr 0 &
0 & 0 & 0 & 0 & 1 \cr -1 & 0 & 0 & 0 & 0 & 0\cr}. }
Knowing this monodromy matrix $a$ and the matrix $A$ from
\monodsym, one can compute $m_{B}$ in \mtp\ using the relation
\monrel.

Now knowing $m_{B}$ and $T$ and $B$ from \monodsym, one can
compute $t$ and $b$ in the Picard-Fuchs basis. They are given
below
\eqn\monodpict{ t= \pmatrix{2 & -1 & 0 & 0 & 0 & 0\cr 1 & 0 & 0 &
0 & 0 & 0\cr -1 & 1 & 1 & 0 & 0 & 0 \cr -2 & 2 & 0 & 1 & 0 & 0 \cr
2 & -2 & 0 & 0 & 1 & 0 \cr 1 & -1 & 0 & 0 & 0 & 1\cr}, }

\eqn\monodpicb{ b= \pmatrix{1 & 0 & 0 & 0 & 0 & 0\cr 1 & 0 & 0 & 0
& -1 & 1\cr -1 & 1 & 1 & 0 & 1 & -1 \cr 0 & 0 & 1 & 0 & 2 & -2 \cr
0 & 0 & -1 & 1 & -1 & 2 \cr 0 & 0 & 0 & 0 & 0 & 1\cr}. }

Next we check explicitly that \monodpict\ is indeed the monodromy
around the conifold point. To this end, following the analysis of
\cofkm\ for the hypersurface in $WP^{8}_{1,1,2,2,2},$ let us
rewrite the periods in the following form
\eqn\perevenodd{\eqalign{ & w_{2j}(\psi,\phi)=-{1\over 6
\pi^3}\sum_{r=1}^{6}(-1)^{r}\sin^{2}({\pi r\over 6})\sin({\pi
r\over 2}) \alpha^{2j r}\xi_{2j}^{r},\cr &
~~~~~~\xi_{2j}^{r}=\sum_{n=1}^{\infty}{\Gamma^{3}(n+{r\over
6})\Gamma(3(n+{r\over 6}))\over \Gamma(6(n+{r\over
6}))}(-1)^{n}(12\psi)^{6n+r}u_{-(n+{r\over 6})}(\phi);\cr &
w_{2j+1}(\psi,\phi)=-{1\over 6
\pi^3}\sum_{r=1}^{6}(-1)^{r}\sin^{2}({\pi r\over 6})\sin({\pi
r\over 2}) \alpha^{(2j+1)r}\xi_{2j+1}^{r},\cr &
~~~~~~\xi_{2j+1}^{r}=\sum_{n=1}^{\infty}{\Gamma^{3}(n+{r\over
6})\Gamma(3(n+{r\over 6}))\over \Gamma(6(n+{r\over
6}))}(12\psi)^{6n+r}u_{-(n+{r\over 6})}(-\phi); }}
where $j=0,1,2$ and $\alpha=\exp({\pi i\over 6})$. Next we follow the analysis
of \cogp\ and \cofkm\ and find
$${d^{2}w_{i}(\psi,0)\over d\psi^2}\sim {\rm const} {c_{i}\over 1-864\psi^6}.$$
for $864\psi^{6}$ in vicinity of $1$ and $c_{i}=(1,1,-1,-2,2,1)$,
for $i=0,..,5$. Here one gets this using the Stirling formula for
the expansion of $\Gamma$-functions and the following result of
\cofkm
\eqn\ununol{
u_{\nu}(0)={2^{\nu}\sqrt{\pi}\exp({\pi i \nu\over 2})
\over \Gamma(1+{\nu\over 2})\Gamma({1\over 2}-{\nu\over 2})}.
}

The resulting monodromy $t$ around the conifold point is given by
\eqn\wj{
w_{j}\rightarrow w_{j}+c_{j}(w_{0}-w_{1}),
}
which exactly coincides with the matrix \monodpict.

\listrefs
\end